\documentclass[]{article}

\usepackage{graphicx}
\usepackage{amssymb}

\newtheorem{theorem}{Theorem}

\newtheorem{definition}[theorem]{Definition}
\newtheorem{example}[theorem]{Example}

\newtheorem{proposition}[theorem]{Proposition}

\newenvironment{proof}[1][Proof]{\noindent\textbf{#1.} }{\ \rule{0.5em}{0.5em}}

\newcommand\JMP{\textit{J. Math. Phys.} }

\begin{document}

\title{Geometric description of lightlike foliations by an
observer in general relativity}

\author{V. J. Bol\'os \\{\small Dpto. Matem\'aticas, Facultad de Ciencias, Universidad de
Extremadura.}\\ {\small Avda. de Elvas s/n. 06071--Badajoz,
Spain.}\\ {\small e-mail\textup{: \texttt{vjbolos@unex.es}}}}

\date{January 31, 2005}

\maketitle

\begin{abstract}
We introduce new concepts and properties of lightlike
distributions and foliations (of dimension and co-dimension 1) in
a space-time manifold of dimension $n$, from a purely geometric
point of view. Given an observer and a lightlike distribution
$\Omega $ of dimension or co-dimension 1, its lightlike direction
is broken down into two vector fields: a timelike vector field $U$
representing the observer and a spacelike vector field $S$
representing the relative direction of propagation of $\Omega $
for this observer. A new distribution $\Omega _U^-$ is defined,
with the opposite relative direction of propagation for the
observer $U$. If both distributions $\Omega $ and $\Omega _U^-$
are integrable, the pair $\left\{ \Omega ,\Omega _U^-\right\} $
represents the wave fronts of a stationary wave for the observer
$U$. However, we show in an example that the integrability of
$\Omega $ does not imply the integrability of $\Omega _U^-$.
\end{abstract}

\section{\label{sec1} Introduction}

It is known that foliations theory is an appropriate framework to
study wave fronts \cite{Sour97}. For example, in symplectic
mechanics, the evolution of a dynamical system is studied by the
leaves of a foliation associated to a symplectic manifold $\left(
\mathcal{U},\sigma \right) $: this dynamical system has a space of
evolution with the structure of a presymplectic manifold $\left(
\mathcal{V},\sigma_{\mathcal{V}}\right) $, whose characteristic
foliation, $ker\sigma_{\mathcal{V}_{\flat }}$, describes
\cite{Sour97} the evolution of $\left( \mathcal{U},\sigma \right)
$. In a particular case, in Special Relativity, for any observer
the evolution of a free massless elementary particle is described
by 2-planes moving in the normal direction at the speed of light
\cite{Sour97} (i.e. lightlike moving wave fronts). This result
that holds in the Minkowski space-time is not trivially
generalized to General Relativity \cite{LiOl95}.

In this paper we are only interested in the dynamics of the wave
fronts as leaves of a foliation. So, we introduce new concepts and
properties of lightlike distributions and foliations (of dimension
and co-dimension 1) in order to improve this description of the
wave fronts, from a purely geometric point of view.

In Sec. \ref{sec2}, \ref{sec3}, \ref{sec4} we work with
distributions to make the study as general as possible, with the
aim of making the results applicable to a wider range of
situations. In Sec. \ref{sec2}, given an observer and a lightlike
distribution $\Omega $, its lightlike direction is broken down
into two vector fields: a timelike vector field $U$ representing
the observer and a spacelike vector field $S$ representing the
relative direction of propagation of $\Omega $ for this observer.
In Sec. \ref{sec3} we study the change of observer, and in Sec.
\ref{sec4} we define a new distribution $\Omega _U^-$ with the
opposite relative direction of propagation for the observer $U$.
In Sec. \ref{sec5} we give more properties about the change of
observer, and in Sec. \ref{sec6} we study the case that both
distributions $\Omega $ and $\Omega _U^-$ are integrable. However,
we show in Example \ref{example1} that the integrability of
$\Omega $ does not imply the integrability of $\Omega _U^-$. In
Sec. \ref{sec7} we give some physical interpretations and
applications of some obtained results. For example, if $\Omega $
and $\Omega _U^-$ are integrable, the pair $\left\{ \Omega ,\Omega
_U^-\right\} $ represents a stationary wave for the observer $U$.
Finally, we apply some results to give an interpretation of the
light aberration in General Relativity.

We will work on an $n$-dimensional space-time manifold $M$ ($n>2$)
with metric $g$ given by $ds^2=g_{ij}dx^idx^j$ and $c=1$. A list
of vector fields inside $span\left( \right)$ will denote the
subbundle generated by these vector fields (called {\it
distribution}). Usually, a distribution of dimension $p$ is called
a $p$-distribution. All adapted bases of distributions are local.
A distribution that has an integral submanifold (leaf) in every
point is a {\it foliation}. We will say that a future pointing
timelike unitary vector field is an observer (identifying it with
the $4$-velocities of a congruence of timelike world lines), and
usually we will denote it by $U$. All the causal vectors
(lightlike or timelike) will be considered future-pointing. We
will denote by $\Omega ^{\bot }$ the orthogonal distribution of
$\Omega $. If $v$ is a vector, $v^{\bot }$ denotes the orthogonal
subspace of $v$. Moreover, if $v$ is a spacelike vector, $\| v\| $
denotes the module of $v$.

\section{\label{sec2} Introducing the concept of ``basis associated to an observer"}

A lightlike 1-distribution is a field of 1-dimensional lightlike
subspaces and a lightlike $(n-1)$-distribution is a
$(n-1)$-dimensional distribution of subspaces whose normal is a
lightlike vector field, which must therefore be contained in the
distribution itself. So, both 1-distributions and
$(n-1)$-distributions are completely determined by a lightlike
vector field $N$ representing the ``lightlike direction'' of the
distribution (up to scale factors). Moreover, two distributions
$\Lambda $ and $\Omega $ (of dimension and co-dimension 1,
respectively) are orthogonal if and only if they have the same
lightlike direction. In this case, the 1-distribution is contained
in the $(n-1)$-distribution: $\Lambda \bot \Omega
\Longleftrightarrow N\in \Lambda \subset \Omega $.

In each case (1-distribution and $(n-1)$-distribution) this
lightlike direction can always be expressed as the sum of an
observer $U$ and an orthogonal unitary spacelike vector field $S$
which represents the relative direction of $N$ for this observer.
So, a basis of a lightlike 1-distribution $\Lambda $ is given by
\begin{equation}
\label{f1} \left\{ S+U\right\}
\end{equation}
where $S\in U^{\bot }$ and $\| S\| =1$. We will say that
(\ref{f1}) is the {\it $U$-basis of $\Lambda $} (also known as
{\it basis of $\Lambda $ associated to $U$}), since it is
completely determined by $U$. Moreover, we will say that $S$ is
the {\it direction of the relative velocity of $\Lambda $ observed
by $U$}.

\begin{example}
In the Minkowski space-time (with $n=4$), expressing the metric
$g$ in rectangular coordinates, we consider the 1-distribution
$\Lambda $ generated by
\begin{equation}
\label{f2} \left\{ \frac{\partial }{\partial z}+\frac{\partial
}{\partial t}\right\} .
\end{equation}
If we consider the observer $U=\frac{\partial }{\partial t}$, then
(\ref{f2}) is the $U$-basis of $\Lambda $ of the form $\left\{
S+U\right\} $ where $S=\frac{\partial }{\partial z}$ is a unitary
spacelike vector field orthogonal to $U$. But if we consider the
observer $U'=\sqrt{2}\left( \frac{\partial }{\partial
t}+\frac{1}{\sqrt{2}}\frac{\partial }{\partial x}\right) $, then
(\ref{f2}) is not the $U'$-basis of $\Lambda $. It is given by
$\left\{ S'+U'\right\} $ where $S'=\sqrt{2}\left( \frac{\partial
}{\partial z}-\frac{1}{\sqrt{2}}\frac{\partial }{\partial
x}\right) $ is a unitary spacelike vector field orthogonal to
$U'$. Logically, $S+U$ and $S'+U'$ represents the same lightlike
direction (i.e. they are proportional).
\end{example}

On the other hand, a lightlike $(n-1)$-distribution $\Omega $ can
therefore only contain $(n-2)$ independent spacelike vector fields
since timelike vectors cannot be orthogonal to lightlike vectors
by simple special relativistic considerations. Given an observer
$U$, a basis of $\Omega $ can therefore be chosen to be $S+U$ (a
representative of the lightlike direction of $\Omega $) and
$(n-2)$ independent spacelike unitary vector fields, $X_1,\ldots
,X_{n-2}$:
\begin{equation}
\label{f3} \left\{ X_1,\ldots ,X_{n-2},S+U\right\}
\end{equation}
such that $X_1,\ldots ,X_{n-2},S$ are an orthonormal basis of
$U^{\bot }$. So $X_1,\ldots ,X_{n-2}$ form an orthonormal basis of
$\Omega \cap U^{\bot }$ and $\left\{ X_1,\ldots
,X_{n-2},S,U\right\} $ form an orthonormal basis of $TM$ (called
tetrad in the case $n=4$, see \cite{Feli90}). We will say that
(\ref{f3}) is a {\it $U$-basis of $\Omega $} (also known as a {\it
basis of $\Omega $ associated to $U$}). Moreover, we will say that
$S$ is the {\it direction of the relative velocity of $\Omega $
observed by $U$}. $X_1,\ldots ,X_{n-2}$ depends on $U$ but they
are not completely determined (we can make infinite choices, since
they form an orthonormal basis of $\Omega \cap U^{\bot }$, i.e.
they complete with $S$ an orthonormal basis of $U^{\bot }$).

\begin{example}
In the Minkowski space-time, expressing the metric $g$ in
rectangular coordinates, we consider the 3-foliation $\Omega $
generated by
\begin{equation}
\label{f4} \left\{ \frac{\partial }{\partial x},\frac{\partial
}{\partial y},\frac{\partial }{\partial z}+\frac{\partial
}{\partial t}\right\} .
\end{equation}
If we consider the observer $U=\frac{\partial }{\partial t}$, then
(\ref{f4}) is a $U$-basis of $\Omega $ of the form $\left\{
X_1,X_2,S+U\right\} $ where $X_1 =\frac{\partial }{\partial x}$,
$X_2 =\frac{\partial }{\partial y}$, $S =\frac{\partial }{\partial
z}$ are an orthonormal basis of $U^{\bot }$. But if we consider
the observer $U'=\sqrt{2}\left( \frac{\partial }{\partial
t}+\frac{1}{\sqrt{2}}\frac{\partial }{\partial x}\right) $, then
(\ref{f4}) is not a $U'$-basis of $\Omega $. It is given by
$\left\{ X_1',X_2',S'+U'\right\} $ where $X_1' =\frac{\partial
}{\partial x}+\frac{1}{\sqrt{2}}\left( \frac{\partial }{\partial
z}+\frac{\partial }{\partial t}\right) $, $X_2' =\frac{\partial
}{\partial y}$, $S' =\sqrt{2}\left( \frac{\partial }{\partial
z}-\frac{1}{\sqrt{2}}\frac{\partial} {\partial x}\right) $ are an
orthonormal basis of $\left( U'\right) ^{\bot }$. It is important
to remark that $S'$ is completely determined by $U'$, but we can
choose $X_1'$, $X_2'$ to complete this orthonormal basis.
Logically, $S+U$ and $S'+U'$ represent the same lightlike
direction (i.e. they are proportional), and the subspace
$span\left( X_1,X_2\right) $ is not the same as the subspace
$span\left( X_1',X_2'\right) $.
\end{example}

So, the vector fields of a basis associated to an observer
describe the distribution from the point of view of this observer.

\section{\label{sec3} Change of observer}

With this notation, given a lightlike distribution (of dimension
or co-dimension 1) and two different observers $U$, $U'$, a change
of observer is a change from a $U$-basis to a $U'$-basis. Now we
are going to study this change: $U'$ can be always written in the
form
\begin{equation}
\label{f5}
U'=\gamma \left( U+vX\right)
\end{equation}
where $X\in U^{\bot }$, $\| X\| =1$, $v$ is a differentiable
function such that $0<v<1$ and $\gamma =\frac{1}{\sqrt{1-v^2}}$.
It is easy to prove that this decomposition is unique. We will say
that $v$ is the {\it module of the relative velocity of $U'$
observed by $U$} and $X$ is the {\it direction of the relative
velocity of $U'$ observed by $U$} \cite{Bolo03}. So, $\gamma $ is
the {\it gamma factor} corresponding to the velocity $v$.

Let $\left\{ S+U\right\} $ and $\left\{ S'+U'\right\} $ be the
$U$-basis and the $U'$-basis, respectively, of the same lightlike
1-distribution $\Lambda $. Then, it is easy to prove that
\begin{equation}
\label{f1.12} S'=\frac{1}{\gamma \left( 1-vg\left( X,S\right)
\right) }\left( S+U\right) -\gamma \left( U+vX\right) .
\end{equation}
The fact that $S'$ is different from $S$ is the {\it aberration}
effect.

In the same way, let $\{X_1,\ldots ,X_{n-2},S+U\}$ and
$\{X_1',\ldots ,X_{n-2}',S'+U'\}$ be a $U$-basis and a $U'$-basis,
respectively, of the same lightlike $(n-1)$-distribution $\Omega
$. Then, $S'$ is given by (\ref{f1.12}) too and, by direct
calculus (orthonormalization of $\left\{ X_1,\ldots
,X_{n-2},S',U'\right\} $), we obtain that a choice of $X_i'$ can
be
\begin{eqnarray*}
X_i' =X_i+\frac{vg\left( X,X_i\right) }{1-vg(X,S)}\left(
S+U\right)
\end{eqnarray*}
for $i=1,\ldots ,n-2$.

\section{\label{sec4} Equivalence relations between distributions and observers}

In this Section, we are going to define equivalence relations
between distributions (see Definition \ref{def1}) and between
observers (see Definition \ref{def2}).

\begin{definition}
\label{def1} Given an observer $U$, let $\Omega $ and $\Omega '$
be two lightlike $p$-distributions of the same dimension
($p=1,n-1$) with lightlike directions $S+U$ and $S'+U$,
respectively. We will say that these $p$-distributions are {\it
equal up to orientations} for the observer $U$, if and only if
$S'=\pm S$. It will be denoted by $\Omega \stackrel{U}{=} \Omega
'$ u.t.o.
\end{definition}

In this case, the relative directions of the lightlike directions
of these $p$-distributions are the same or the opposite for this
observer.

Given $\Omega $ a lightlike $p$-distribution ($p=1,n-1$) and $U$
an observer, we can build another lightlike $p$-distribution with
the opposite relative direction of propagation for this observer,
and it will be denoted by $\Omega _U^-$. So, if $p=1$ and $\left\{
S+U\right\} $ is the $U$-basis of $\Omega $, then $\left\{
-S+U\right\} $ is the $U$-basis of $\Omega _U^-$. On the other
hand, if $p=n-1$ and $\left\{ X_1,\ldots ,X_{n-2},S+U\right\} $ is
a $U$-basis of $\Omega $, then $\left\{ X_1,\ldots
,X_{n-2},-S+U\right\} $ is a $U$-basis of $\Omega _U^-$. In both
cases ($p=1,n-1$) we have that $\Omega _U^-$ is well defined and
obviously $\left( \Omega _U^-\right) _U^-=\Omega _{UU}^{--}=\Omega
$.

It is not difficult to prove that $\Omega \stackrel{U}{=}\Omega '$
u.t.o. is equivalent to $\Omega ^{\bot }\stackrel{U}{=}\Omega
'{}^{\bot }$ u.t.o., and it is clear that
\begin{equation} \label{fort} \left(
\Omega _U^-\right) ^{\bot }=\left( \Omega ^{\bot }\right) _U^-.
\end{equation}

\begin{figure}
\begin{center}
\includegraphics[width=0.3\textwidth]{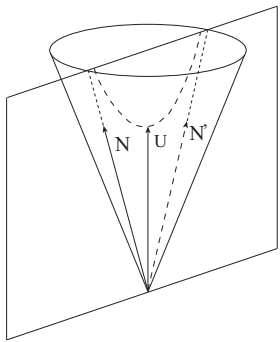}
\end{center}
\caption{\label{fig1} The lightlike directions of $\Omega $ and
$\Omega _U^-$ are represented as $N$ and $N'$, respectively
(lightlike vectors in the future lightcone), and $U$ is
represented as a vector inside the lightcone, since it is
future-pointing timelike unitary}
\end{figure}

\begin{figure}
\begin{center}
\includegraphics[width=0.4\textwidth]{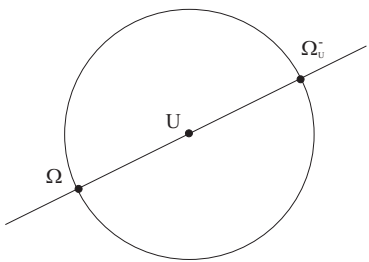}
\end{center} \caption{\label{fig2} Figure \ref{fig1} could be
represented schematically in a section of the lightcone: lightlike
directions are represented as points of the circle (that
represents the section), observers are represented as points
inside the circle and the 2-plane defined by the lightlike
directions of $\Omega $ and $\Omega _U^-$ is represented as a
straight line joining both lightlike directions. Since a lightlike
direction determines the lightlike $p$-distribution ($p=1,n-1$),
the lightlike direction is represented by the name of the
distribution which determines. So, the lightlike direction of
$\Omega $ is represented as ``$\Omega $'' and the lightlike
direction of $\Omega _U^-$ is represented as ``$\Omega _U^-$''}
\end{figure}

Summing up, given an observer $U$, the relation ``be equal up to
orientations for the observer $U$'' defines a quotient space on
the set of the lightlike distributions of dimension or
co-dimension 1. This quotient space is the framework to study
stationary waves, as we will see in Sec. \ref{sec7.2}. Each
equivalence class is defined by two distributions, $\Omega $ and
$\Omega _U^-$, i.e.:
\begin{equation}
\label{relation} \Omega \stackrel{U}{=}\Omega '\quad u.t.o.\quad
\Longleftrightarrow \quad \Omega '=\Omega \quad or\quad \Omega
'=\Omega _U^-.
\end{equation}
From a geometrical point of view, the lightlike directions of
$\Omega $ and $\Omega _U^-$ are in a $2$-plane which contains the
observer $U$ (see Fig. \ref{fig1} and Fig. \ref{fig2}).

Let us now introduce the concept of ``observers $\Omega
$-related":

\begin{definition}
\label{def2} Given a lightlike distribution $\Omega $ (of
dimension or co-dimension 1) and two observers, $U$ and $U'$, we
will say that these observers are {\it $\Omega $-related} if and
only if $\Omega _U^-=\Omega _{U'}^{-}$. It will be denoted by
$U\stackrel{\Omega }{\approx }U'$.
\end{definition}

\begin{figure}
\begin{center}
\includegraphics[width=0.45\textwidth]{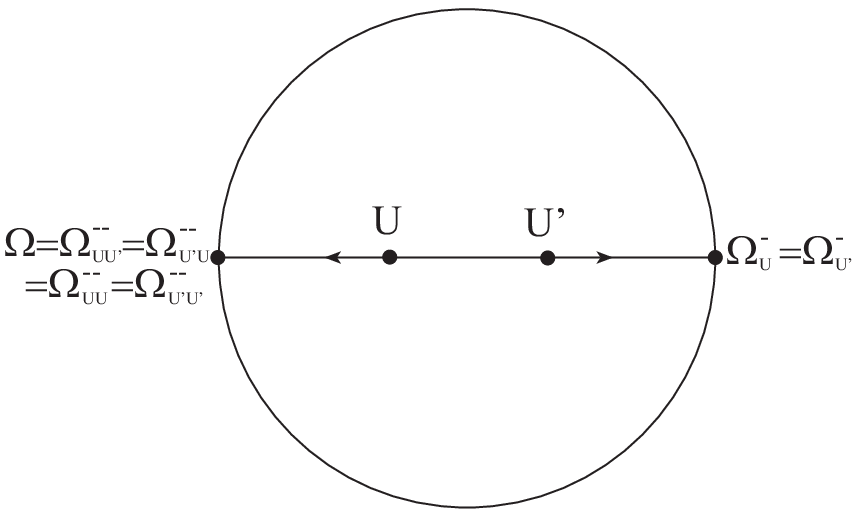}
\end{center} \caption{\label{fig3} Scheme of the properties
concerning to two observers $\Omega $-related}
\end{figure}

Then, two observers are $\Omega $-related if and only if they are
coplanar with the lightlike direction of $\Omega $ (see Fig.
\ref{fig3}). So, if these observers are observing $\Omega $, there
is not any aberration effect between them, as we will see in Sec.
\ref{sec7.1}.

\begin{figure}
\begin{center}
\includegraphics[width=0.35\textwidth]{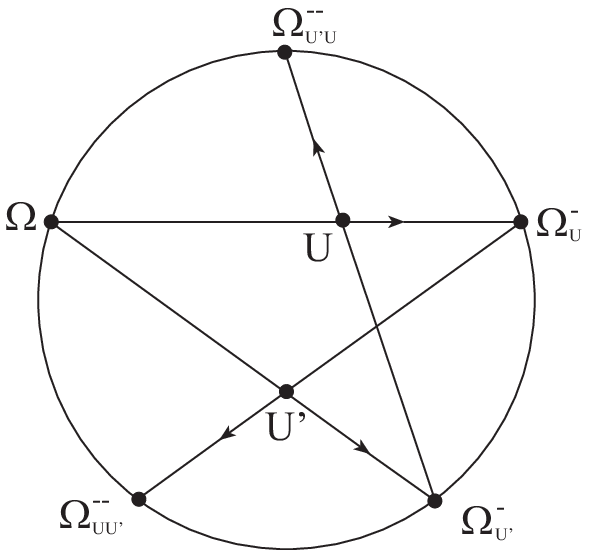}
\end{center} \caption{\label{fig4} Scheme of two observers that are
not $\Omega $-related}
\end{figure}

Properties of observers $\Omega $-related are compiled in the next
theorem:

\begin{theorem}
\label{t0} Given a lightlike $p$-distribution $\Omega $
($p=1,n-1$) and two observers, $U$ and $U'$, the following
properties are equivalent (see Fig. \ref{fig3} and Fig.
\ref{fig4}):

\begin{itemize}

\item[(i)]$U\stackrel{\Omega }{\approx }U'$, i.e. $U$ and $U'$ are
$\Omega$-related.

\item[(ii)]$U\stackrel{\Omega _U^-}{\approx }U'$, i.e. $U$ and
$U'$ are $\Omega _U^-$-related.

\item[(iii)] $\Omega _U^-=\Omega _{U'}^-$.

\item[(iv)] $U,U'$ and the lightlike direction of $\Omega $ are in
the same $2$-plane.

\item[(v)] $\Omega \cap U^{\bot }=\Omega \cap \left( U'\right)
^{\bot }$.

\item[(vi)] $U'\in \left( \Omega \cap U^{\bot }\right) ^{\bot }$.

\item[(vii)] $\Omega _{UU'}^{--}=\Omega _{U'U}^{--}=\Omega $.

\end{itemize}

\end{theorem}

\begin{proof}

All the equivalences can be easily proved, except for the property
\begin{equation} \label{ft1}
\Lambda _{UU'}^{--}=\Lambda _{U'U}^{--}\longrightarrow
U\stackrel{\Omega }{\approx }U',
\end{equation}
where $\Lambda $ is a lightlike $1$-distribution (the case of
dimension $p=n-1$ can be proved taking into account (\ref{fort})
and (\ref{ft1})).

Let $\left\{ S+U\right\} $ be the $U$-basis of $\Lambda $ and let
$\left\{ X_{1},\ldots ,X_{n-2},S,U\right\} $ be an orthonormal
basis.

\begin{itemize}
\item  We are going to build the $U'$-basis of $\Lambda
_{UU'}^{--}$:

We have that $\left\{ -S+U\right\} $ is the $U$-basis of $\Lambda
_{U}^{-}$. If $\left\{ -Y'+U'\right\} $ is the $U'$-basis of
$\Lambda _{U}^{-}$, then $\left\{ Y'+U'\right\} $ is the
$U'$-basis of $\Lambda _{UU'}^{--}$. We can write $U'=\gamma
\left( U+vX\right) $ (see (\ref{f5})), and by using algebraic
manipulations, we have that
\[
Y'+U' = 2v\gamma g\left( X,X_1\right) X_{1}
\]
\[
+\ldots +2v\gamma g\left( X,X_{n-2}\right) X_{n-2}
\]
\[
+\left( 2v\gamma g\left( X,S\right) +\frac{1}{\gamma \left(
1+vg\left( X,S\right) \right) }\right) S
\]
\begin{equation} \label{f1.35}
+\left(2\gamma -\frac{1}{\gamma \left( 1+vg\left( X,S\right)
\right) }\right) U.
\end{equation}

\item We are going to build the $U$-basis of $\Lambda
_{U'U}^{--}$:

Let $\left\{ -S'+U'\right\} $ be the $U'$-basis of $\Lambda
_{U'}^{-}$ and let $\left\{ X'_{1},\ldots ,X'_{n-2},S',U'\right\}
$ be an orthonormal basis. If $\left\{ -Y+U\right\} $ is the
$U$-basis of $\Lambda _{U'}^{-}$, then $\left\{ Y+U\right\} $ is
the $U$-basis of $\Lambda _{U'U}^{--}$. Taking into account
expression (\ref{f5}) and by using algebraic manipulations, we
have that
\[
Y+U = -\frac{2v}{2-\alpha }g\left( X,X_1\right) X_{1}
\]
\[
-\ldots -\frac{2v}{2-\alpha }g\left( X,X_{n-2}\right) X_{n-2}
\]
\begin{equation} \label{f1.36}
+\left( \frac{\alpha -2vg\left( X,S\right) }{2-\alpha }\right)
S+U,
\end{equation}
where $\alpha = \frac{1}{\gamma ^2\left( 1-vg\left( X,S\right)
\right) }$.
\end{itemize}

By the hypothesis, the vector fields $Y'+U'$ and $Y+U$ are
proportional. So, identifying the coefficients of $U$ in
(\ref{f1.35}) and (\ref{f1.36}) we obtain that
\begin{equation} \label{f365}
Y'+U'=\left( 2\gamma -\frac{1}{\gamma \left( 1+vg\left( X,S\right)
\right) }\right) \left( Y+U\right) .
\end{equation}

If we suppose $g\left( X,S\right) \neq \pm 1$ (i.e. $\left|
g\left( X,S\right) \right| <1$), then $g\left( X,X_i\right) \neq
0$ for all $i=1,\ldots ,n-2$. So, identifying the coefficients of
$X_{i}$ in (\ref{f1.35}) and (\ref{f1.36}), taking into account
(\ref{f365}), we obtain that
\begin{equation} \label{f1.34}
2v\gamma =\left( 2\gamma -\frac{1}{\gamma \left( 1+vg\left(
X,S\right) \right) }\right) \left( \frac{-2v}{2-\alpha }\right) .
\end{equation}
Solving $g\left( X,S\right) $ in (\ref{f1.34}), with $v\neq 0$
\[
g\left( X,S\right) =\pm \sqrt{\frac{v^{2}+1}{2v^{2}}},
\]
but $v^{2}<1$, so $\frac{v^{2}+1}{2v^{2}}>1$ and then there is not
any solution with $\left| g\left( X,S\right) \right| <1$.

On the other hand, it can be easily proved that if we suppose
$g\left( X,S\right) =\pm 1$ (i.e. $X=\pm S$), then $Y'+U'$ and
$Y+U$ are always proportional. In this case, taking into account
expression (\ref{f5}) we have that $U'$ can be expressed as a
linear combination of $U$ and $S+U$. So, $U\stackrel{\Omega
}{\approx }U'$ (by (iv)).

\end{proof}

\section{\label{sec5} Equality between associated bases}

We can wonder if given a lightlike distribution (of dimension or
co-dimension 1) and given two different observers, $U$ and $U'$,
it is possible for a $U$-basis to coincide with a $U'$-basis of
this distribution. First, we are going to study it in the case of
dimension 1:

Let $\left\{ S+U\right\} $ be the $U$-basis of a lightlike
1-distribution $\Lambda $. We are going to build all the different
observers $U'$ with $U'$-basis of $\Lambda $ equal to the
$U$-basis of $\Lambda $. Taking into account (\ref{f5}) we have to
find $X$ and $v$ such that
\begin{equation}
\label{ff12} g\left( X,S\right) =\frac{1-\sqrt{1-v^2}}{v}.
\end{equation}
The function $\frac{1-\sqrt{1-v^2}}{v}:\left] 0,1\right[
\rightarrow \left] 0,1\right[ $ is bijective. So, given a
spacelike unitary vector field $X$ orthogonal to $U$ such that
$0<g\left( X,S\right) <1$, there exists a unique $v$ such that
(\ref{ff12}) is satisfied. Then $U'$ is given by (\ref{f5}). So,
we have a different $U'$ for each different $X$ which satisfies
all these conditions.

\begin{example}
In the Minkowski space-time, expressing the metric $g$ in
rectangular coordinates, we consider the lightlike 1-distribution
$\Lambda $ generated by
\begin{equation}
\label{ff13} \left\{ \frac{\partial }{\partial z}+\frac{\partial
}{\partial t}\right\} .
\end{equation}
If we consider $U=\frac{\partial }{\partial t}$ as an observer,
then (\ref{ff13}) is the $U$-basis of $\Lambda $, of the form
$\left\{ S+U\right\} $ where $S=\frac{\partial }{\partial z}$. If
we consider, for instance, the spacelike unitary vector field
$X=\frac{1}{\sqrt{2}}\left( \frac{\partial }{\partial
z}+\frac{\partial }{\partial x}\right) $ orthogonal to $U$, we
have that $g\left( X,S\right) =\frac{1}{\sqrt{2}}\in \left]
0,1\right[ $ and then, we have that $v=\frac{2\sqrt{2}}{3}$ from
(\ref{ff12}). So, given the observer
\[
U'=\frac{1}{\sqrt{1-v^2}}\left( U+vX\right) =3\frac{\partial
}{\partial t}+2\left( \frac{\partial }{\partial z}+\frac{\partial
}{\partial x}\right) ,
\]
we have that the $U'$-basis of $\Lambda $ is the same as the
$U$-basis of $\Lambda $.
\end{example}

In the case of co-dimension 1, we need the concept of ``observers
$\Omega $-related", introduced in Sec. \ref{sec4}. Given $\Omega $
a lightlike $(n-1)$-distribution and given $U$ an observer with
$\left\{ X_1,\ldots ,X_{n-2},S+U\right\} $ a $U$-basis of $\Omega
$, we want to find an observer $U'$ with $U'$-basis of $\Omega $
equal to $\left\{ X_1,\ldots ,X_{n-2},S+U\right\} $. If a
$U'$-basis of $\Omega $ is given by $\left\{ X'_1,\ldots
,X'_{n-2},S'+U'\right\} $, we must impose two conditions:

\begin{itemize}

\item $S+U=S'+U'$

\item Since we have a certain freedom in the choice of
$X'_1,\ldots ,X'_{n-2}$, we have to impose only that $span\left(
X_1,\ldots ,X_{n-2}\right)=span\left( X'_1,\ldots ,X'_{n-2}\right)
$, i.e. $\Omega \cap U^{\bot }=\Omega \cap \left( U'\right) ^{\bot
}$.

\end{itemize}

The first condition is equivalent (according to the case of
dimension 1) to the existence of a spacelike unitary vector field
$X$ orthogonal to $U$ such that $0<g\left( X,S\right) <1$.

The second condition is equivalent (according to Theorem \ref{t0}
(v)) to $U\stackrel{\Omega }{\approx }U'$. Then, $S,U,U'$ are in
the same $(n-2)$-plane, i.e. $U'$ can be expressed as a linear
combination of $S$ and $U$. Since $U'=\gamma \left( U+vX\right) $,
we obtain that $X$ can be also expressed as a linear combination
of $S$ and $U$. But $X$ is orthogonal to $U$ and then, the unique
possibility is $X=\pm S$ (since they are unitary vector fields).
So, $g\left( X,S\right) =\pm 1$, and this is not possible,
according to the first condition.

So, given two different observers, $U$ and $U'$, and a lightlike
$(n-1)$-distribution $\Omega $, we have that a $U$-basis of
$\Omega $ is always different from any $U'$-basis of $\Omega $.

\section{\label{sec6} Conditions of integrability}

In this paper, we have studied some properties of lightlike
$(n-1)$-distributions in general, which only in some cases are
foliations (however, since every vector field is integrable, given
any 1-distribution it is automatically a 1-foliation). Now we are
going to study when we can integrate a lightlike
$(n-1)$-distribution (see Theorem \ref{t1}).

It is well known that for any given observer $U$, its local rest
space $U^{\bot } $ is a foliation if and only if $U$ is
synchronizable \cite{Sach77}. Taking this into account and given a
lightlike $(n-1)$-distribution $\Omega $, we are going to study
sufficient conditions for the integrability of the spacelike
$(n-2)$-distribution $\Omega \cap U^{\bot }$ in next proposition:

\begin{proposition}
\label{prop1} Let $\Omega $ be a lightlike $(n-1)$-distribution,

\begin{itemize}
\item[(i)] if $\Omega $ is a foliation, then $\Omega \cap U^{\bot
}$ is also a foliation for any synchronizable observer $U$.

\item[(ii)] if there exists a synchronizable observer $U$ such
that $\left( \Omega \cap U^{\bot }\right) \oplus span\left(
U\right) $ is a foliation, then $\Omega \cap U^{\bot }$ is a
foliation.
\end{itemize}
\end{proposition}

\begin{proof}

\begin{itemize}

\item[(i)]

given $U$ a synchronizable observer, let $\left\{ X_{1},\ldots
,X_{n-2},S+U\right\} $ be a $U$-basis of $\Omega $.

Since $\Omega $ is a foliation, we have that $\left[
X_{i},X_{j}\right] \in \Omega $ for all $i,j=1,\ldots ,n-2$.

Moreover, $U^{\bot }$ is also a foliation (because $U$ is
synchronizable), so $\left[ X_{i},X_{j}\right] \in U^{\bot }$ for
all $i,j=1,\ldots ,n-2$, and then
\[
\left[ X_{i},X_{j}\right] \in \Omega \cap U^{\bot },
\]
for all $i,j=1,\ldots ,n-2$. So, $\Omega \cap U^{\bot }$ is a
foliation.

\item[(ii)]

let $\left\{ X_{1},\ldots ,X_{n-2},S+U\right\} $ be a $U$-basis of
$\Omega $.

Since $\left( \Omega \cap U^{\bot }\right) \oplus span\left(
U\right) $ is a foliation, we have that $\left[ X_{i},X_{j}\right]
\in \left( \Omega \cap U^{\bot }\right) \oplus span\left( U\right)
$ for all $i,j=1,\ldots ,n-2$.

Moreover, $U^{\bot }$ is also a foliation, so $\left[
X_{i},X_{j}\right] \in U^{\bot }$ for all $i,j=1,\ldots ,n-2$, and
then
\[
\left[ X_{i},X_{j}\right] \in \left( \left( \Omega \cap U^{\bot
}\right) \oplus span\left( U\right) \right) \cap U^{\bot }=\Omega
\cap U^{\bot },
\]
for all $i,j=1,\ldots ,n-2$. So, $\Omega \cap U^{\bot }$ is a
foliation.

\end{itemize}

\end{proof}

Nevertheless, the integrability of the timelike
$(n-1)$-distribution $\left( \Omega \cap U^{\bot }\right) \oplus
span\left( U\right) $ depends exclusively on the integrability of
$\Omega _U^-$, as we see in next theorem:

\begin{theorem}
\label{t1} Let $\Omega $ be a lightlike $(n-1)$-foliation and $U$
a synchronizable observer. $\Omega _U^-$ is a foliation if and
only if $\left( \Omega \cap U^{\bot }\right) \oplus span\left(
U\right) $, is a foliation.
\end{theorem}

\begin{proof}

Let $\left\{ X_{1},\ldots ,X_{n-2},S+U\right\} $ be a $U$-basis of
$\Omega $.

By Proposition \ref{prop1} (i) we have that $\Omega \cap U^{\bot
}$ is a foliation.

\begin{description}
\item[$\longrightarrow $] Moreover, taking into account that
$\Omega $ and $\Omega _U^-$ are foliations, it can be proved that
$\left[ X_{i},U\right] \in span\left( X_{1},\ldots
,X_{n-2},U\right) = \left( \Omega \cap U^{\bot }\right) \oplus
span\left( U\right)$ for all $i=1,\ldots ,n-2$, and so $\left(
\Omega \cap U^{\bot }\right) \oplus span\left( U\right) $ is a
foliation.

\item[$\longleftarrow $] Taking into account that $\Omega $ and
$\left( \Omega \cap U^{\bot }\right) \oplus span\left( U\right) $
are foliations, it can be proved that $\left[ X_{i},-S+U\right]
\in \Omega _U^-$ for all $i=1,\ldots ,n-2$, and so $\Omega _U^-$
is a foliation.
\end{description}

\end{proof}

Theorem \ref{t1} assures that the observation of the foliation
$\Omega _U^-$ is guaranteed only when $\left( \Omega \cap U^{\bot
}\right) \oplus span\left( U\right) $ is a foliation. Note that
Theorem \ref{t1} does not state that if $\Omega $ is a lightlike
$(n-1)$-foliation and $U$ is a synchronizable observer, then
$\Omega _U^-$ is a foliation. Actually we show with an example
that this statement is false:

\begin{example}
\label{example1} In the Minkowski space-time, expressing the
metric $g$ in rectangular coordinates, we consider the following
vector fields:
\begin{eqnarray*}
U &=&\frac{\partial }{\partial t}+a\frac{\partial }{\partial
x}+b\frac{\partial }{\partial y}+c\frac{\partial }{\partial
z}=\left( 1,a,b,c\right)
\\
X_{1} &=&a\frac{\partial }{\partial t}+\frac{\partial }{\partial
x}=\left(
a,1,0,0\right)  \\
X_{2} &=&b\frac{\partial }{\partial t}+\frac{\partial }{\partial
y}=\left(
b,0,1,0\right)  \\
S &=&\frac{c}{d}\left( \frac{\partial }{\partial
t}+a\frac{\partial }{\partial x}+b\frac{\partial }{\partial
y}+\frac{\alpha ^2}{c}\frac{\partial }{\partial z}\right)
=\frac{c}{d}\left( 1,a,b,\frac{\alpha ^2}{c}\right)
\end{eqnarray*}
where $a,b,c,d$ are smooth functions such that $c^2<1-\left(
a^2+b^2\right) =d^2$, $U$ is an observer and $\left\{
X_1,X_2,S\right\} $ are spacelike vector fields defining an
orthogonal basis of $U^{\bot }$. We consider $\Omega $ the
3-distribution generated by $\left\{ X_1,X_2,S+U\right\} $ where
$S+U=\left( 1+\frac{c}{d}\right) \left( 1,a,b,d\right) $. Stating
$b=0$, defining $d$ by the equation
$\frac{x}{d\sqrt{1-d^2}}-\frac{z}{\left( f+d^3\right) ^{2/3}}=2$,
and taking
\[
f=\left( \frac{1}{4}\right) ^{3/2}-\left( \frac{1}{2}\right)
^{3/2},\quad c=\left( f+d^3\right) ^{1/3},\quad a^2=1-d^2,
\]
we obtain that in a neighborhood of $W:=\{\left( t,x,y,z\right)
:x=1,z=0\}\subset M$, $U$ is synchronizable, $\Omega $ is a
foliation and $\Omega _U^-$ is not a foliation.
\end{example}

Moreover, Example \ref{example1} ensures that conditions (i) and
(ii) of Proposition \ref{prop1} are not necessary conditions for
the integrability of $\Omega \cap U^{\bot }$:
\begin{itemize}
\item[(i)] $\Omega _U^-$ is not a foliation, but $\Omega _U^-\cap
U^{\bot }$ is a foliation.

\item[(ii)] $U$ is synchronizable and $\left( \Omega \cap U^{\bot
}\right) \oplus span\left( U\right) $ is not a foliation, but
$\Omega \cap U^{\bot }$ is a foliation.
\end{itemize}

\section{\label{sec7} Some physical interpretations and applications}

According to Sec. \ref{sec1}, a congruence of lightlike rays (rays
of light for example) are, in fact, the leaves of a lightlike
1-foliation $\Lambda $. On the other hand, a congruence of
lightlike moving wave fronts are the leaves of a lightlike $\left(
n-1\right) $-foliation $\Omega $. Precisely, given an observer
$U$, the leaves of $\Omega \cap U^{\bot }$ can be interpreted as
the spacelike wave fronts ($\left( n-2\right) $-dimensional
without movement) for this observer \cite{LiOl95}. But neither all
lightlike 1-foliations can be interpreted as a congruence of
lightlike rays nor all lightlike $\left( n-1\right) $-foliations
can be interpreted as a congruence of lightlike moving wave
fronts. For example, $\Lambda $ and $\Omega $ should be totally
geodesic foliations. Moreover, in the case of dimension $\left(
n-1\right) $, condition (i) of Proposition \ref{prop1} should be a
sufficient and necessary condition: $\Omega $ is integrable if and
only if $\Omega \cap U^{\bot }$ is integrable (given a
synchronizable observer $U$). This fact is reasonable, since the
observation of the wave fronts (as leaves of $\Omega \cap U^{\bot
}$) should imply the existence of the wave fronts (as leaves of
$\Omega $) and viceversa. In this Section, we are going to work in
a 4-dimensional space-time (i.e. $n=4$) with foliations that can
be interpreted in this way (as congruences of lightlike rays or
lightlike moving wave fronts) to make physical applications of
some results given in this paper.

\subsection{\label{sec7.1} Light aberration}

Let $\Omega $ be a lightlike foliation of dimension 1 or 3, and
let $U$, $U'$ be two observers. If $N$ represents the lightlike
direction of $\Omega $, let $S$, $S'$ represent the relative
direction of $N$ for the observers $U$, $U'$, respectively (i.e.
$S+U$ and $S'+U'$ are proportional to $N$). These relative
directions are the spacelike directions of propagation of the
lightlike rays (dimension 1) or the lightlike moving wave fronts
(dimension 3) for these observers, respectively. So, we have that
\begin{itemize}
\item if $U\stackrel{\Omega }{\approx }U'$ (i.e. they are $\Omega
$-related), then $S$ and $S'$ are proportional. In this case, they
represent the same relative direction.

\item if $U$ and $U'$ are not $\Omega $-related, then $S$, $S'$
represent different relative directions. As we said in Sec.
\ref{sec3}, this fact is the light aberration effect.

Taking into account (\ref{f1.12}) we can obtain, for example, the
usual expression for light aberration (see \cite{Syng65}) in a
more general way:
\[
\cos \theta =\frac{\cos \theta '-v}{1-v\cos \theta '},
\]
where $\theta $ is the angle between $-S$ and $X$, (i.e. $\cos
\theta =g\left( X,-S\right) $) and $\theta '$ is the angle between
$-S'$ and the projection of $X$ to $U'^{\bot }$.
\end{itemize}

\subsection{\label{sec7.2} Stationary waves}

It is known that a stationary wave is formed in fact by two
identical waves with opposite relative directions of propagation.
So, a 3-dimensional lightlike stationary wave for an observer $U$
can be represented as a pair of lightlike 3-foliations $\left\{
\Omega ,\Omega _U^-\right\} $ (actually they represent the
3-dimensional wave fronts), since the foliations $\Omega $ and
$\Omega _U^-$ have opposite relative directions of propagation for
the observer $U$. So, a 3-dimensional lightlike stationary wave
for an observer $U$ is represented by an equivalence class of the
relation ``be equal up to orientations for the observer $U$'',
given in Definition \ref{def1} and in expression (\ref{relation}).
In this case, if $U$ and $U'$ are two observers $\Omega $-related,
$\left\{ \Omega ,\Omega _U^-\right\} $ represents a stationary
wave for both observers $U$ and $U'$.

Finally, as we are only interested in lightlike 3-foliations
$\Omega $ with physical interpretation (representing congruences
of lightlike moving wave fronts), condition (i) of Proposition
\ref{prop1} should be a sufficient and necessary condition (this
fact is discussed at the beginning of this Section). So, given $U$
a synchronizable observer, $\Omega _U^-$ should be also a
foliation, since $\Omega \cap U^{\bot }=\Omega _U^-\cap U^{\bot }$
is a foliation. If we take this into account, Example
\ref{example1} does not have any physical interpretation, since
$\Omega $ cannot represent any congruence of lightlike moving wave
fronts. In fact, $\Omega $ is not a totally geodesic foliation.

\end{document}